%% file: main.tex
\documentclass[submission,copyright,creativecommons]{eptcs}
\providecommand{\event}{LFMTP 2020} 
\usepackage{breakurl}             
\usepackage{underscore}           
\usepackage[utf8]{inputenc}

\usepackage{listings}
\input{dkset}
\newcommand{\lst}{\lstinline}

\usepackage[english]{babel}
\usepackage[dvipsnames]{xcolor}
\usepackage{mathtools, amssymb}

\usepackage{multicol}
\usepackage[thmmarks, amsmath]{ntheorem}
\usepackage{bussproofs}

\usepackage{graphicx}
\usepackage{subcaption}

\usepackage{array}

\usepackage{amsfonts}
\usepackage{amsmath}

\usepackage{tikz}
\usetikzlibrary{calc}
\usetikzlibrary{patterns}

\usepackage{stmaryrd}

\makeatletter
\newcommand{\institutionlogo}[1]{\def\@institutionlogo{#1}}
\newcommand{\institution}[1]{\def\@institution{#1}}
\newcommand{\supervisor}[1]{\def\@supervisor{#1}}
\newcommand{\schoolyear}[1]{\def\@schoolyear{#1}}
\newcommand{\school}[1]{\def\@school{#1}}
\newcommand{\cur}[1]{\def\@cur{#1}}

\makeatother

\newcommand{\newZdSbox}[3]{{
\theoremprework{~\\ \textcolor{#3}{\rule{0.6\linewidth}{1pt}}}
\theorempostwork{\hfill \textcolor{#3}{\rule{0.6\linewidth}{1pt}} \\}
\theoremheaderfont{\scshape}
\theoremseparator{ ---}
\theoremstyle{break}
\theorembodyfont{\normalfont}
\newtheorem*{#1}{\textcolor{#3}{#2}}
}}

\newZdSbox{Example}{Example}{ForestGreen}

\newZdSbox{Summary}{Summary}{Plum}

\newZdSbox{Theorem}{Theorem}{Orange}
\newZdSbox{Conjecture}{Conjecture}{Orange}

\newZdSbox{Definition}{Definition}{Blue}

\newZdSbox{Proof}{Proof}{Red}

\newcommand*{\ie}{\textit{i.e. }}

\title{Implementation of Two Layers Type Theory in Dedukti \\and Application to Cubical Type Theory}
\author{Bruno Barras
\institute{Inria, Université Paris-Saclay,\\
ENS Paris-Saclay, CNRS, LSV,\\
91190, Gif-sur-Yvette, France.}
\and
Valentin Maestracci
\institute{Université Paris-Saclay,\\
ENS Paris-Saclay, CNRS, LSV,\\
91190, Gif-sur-Yvette, France.}
}
\def\titlerunning{Two Layers Type Theory in Dedukti}
\def\authorrunning{B. Barras \& V. Maestracci}

\input{rules/operators}






\begin{document}
\maketitle

\begin{abstract}
In this paper, we make a substantial step towards an encoding of
Cubical Type Theory (CTT) in the Dedukti logical
framework. Type-checking CTT expressions features a decision procedure
in a de Morgan algebra that so far could not be expressed by the
rewrite rules of Dedukti. As an alternative, 2 Layer Type Theories are
variants of Martin-L\"of Type Theory where all or part of the
definitional equality can be represented in terms of a so-called
external equality. We propose to split the encoding by giving an
encoding of 2 Layer Type Theories (2LTT) in Dedukti, and a partial
encoding of CTT in 2LTT.
\end{abstract}

\section{Introduction}

\input{intro}

\section{Two Layers Type Theories in Dedukti}

    \subsection{Two Layer Type Theory as a Dedukti theory}
        \input{sub/theory/2ltt}

    \subsection{Translating Two Layer Type Theories}

\input{sub/implem/translation}

    \subsection{Soundness of the encoding}

\input{sub/implem/soundness}

\section{Cubical Type Theory in Dedukti}

        
        \input{sub/explanations/clttare2ltt}


    \subsection{Systems}
        \input{sub/implem/cttsys}

    \subsection{Composition and the example of filling}
        \input{sub/implem/cttfill}

    \subsection{Translation of Cubical expressions}
        \input{sub/implem/cttsoundness}

\section{Conclusion}

\input{concl}


\nocite{*}
\bibliographystyle{eptcs}
\bibliography{stage}
\end{document}

%% file: dkset.tex
\lstdefinelanguage{Dedukti}{
  alsoletter={=->:\#},
  keywords={Type,def,defac,-->,->,=>,:=,:,.,\#SNF,\#NAME,\#PRINT},
  delim=[s][\color{brown}]{[}{]},
  comment=[n]{(;}{;)},
  string=[b]{"},
  stringstyle=\color{orange},
  commentstyle=\color{red},
  showstringspaces=false
}

%% file: rules/operators.tex
\let\th\vdash

\newcommand{\Inst}[2]{#1[#2]}
\newcommand{\Trad}[2]{{\llbracket#2\rrbracket}_{#1}}

\newcommand{\Sum}[3]{\underset{x:#2}{\Sigma^{#1}}(#3)}
\newcommand{\pair}[5]{\text{pair}^{#1}_{[x:#2]#3}(#4,#5)}


%% file: intro.tex
The goal of this paper is to explore the possibility to express
Homotopy Type Theory (HoTT,~\cite{HottBook}) in the
Dedukti~\cite{assaf} logical framework.


Dedukti is a logical framework which main distinctive
feature is the possibility to extend the definitional equality (aka
conversion) with a class of rewrite rules. It is intended to be used
as a ``hub'' for proof systems. Many of the logics implemented by
those systems can be encoded as \emph{Dedukti theories}, and proofs in
those systems can be expressed as Dedukti terms in the corresponding
theory. The point is not just to collect the proofs of many logics,
but rather to make it easier to translate proofs from one system to
the other.

Expressing HoTT as a Dedukti theory is at the same time a problem that challenges the expressiveness
of the rewrite rules accepted by Dedukti, but HoTT is an interesting
formalism on its own. Moreover, in the long term, it would be
interesting to have tools to investigate which proofs can be adapted
from more conventional logics (HOL, set theory) to HoTT and conversely.


The paper is organized as follows. We first give an introduction on
HoTT, Dedukti and encodings of Type Theory in Dedukti. This will
motivate the introduction of Two Level Type Theories (2LTT) as a more
flexible way to deal with type theories having a definitional equality
two complex to be encoded as rewrite rules. In the second section, we
will introduce 2LTTs and their encoding as a Dedukti theory. In the
third section, we will give a partial encoding of Cubical as a 2LTT,
and the corresponding piece of Dedukti theory.

Both encodings have been implemented and can be found here:
\begin{center}\url{https://github.com/valent20000/CTTDedukti}.\end{center}

\subsection{Homotopy Type Theories}


In the broad sense, HoTT is based on an interpretation of
Type Theory (ML, Coq, Agda, NuPRL) where types are topological spaces and
proofs of an equality $x=y$ in type $A$ are continuous paths between
points $x$ and $y$ of space $A$.



This specific interpretation implies the possibility to extend the
theory with new principles. The most famous one is \emph{univalence}
expressing that paths between types are exactly the weak equivalences
(which is a particular case of isomorphism that deals with higher
dimensions). Other extensions include higher inductive types, which
are a generalization of inductive definitions. They are used to define
spaces such as the circle, the torus, suspensions and many others.

Some members of the HoTT family are just regular Type Theory extended
with axioms. The drawback of this approach is that axioms breaks
metatheoretical properties such as canonicity. Worse, some useful
notions such as simplicial sets seems impossible to express by mere
axioms: face map equations require a coherence condition at dimension 2, which in turn requires another coherence condition at higher dimension, etc. This situation is often called ``coherence hell''.

In contrast, several members of the HoTT family, called \emph{cubical} (\cite{CCHM},\cite{RedPRL}), give
a computational interpretation to univalence. In this paper we will focus on the Cubical Type Theory in~\cite{CCHM} (called Cubical from now on) that we will briefly introduce in the next section. We believe that adapting this work to the others cubical theories should be easy.


\subsection{Cubical Type Theory}

\input{sub/theory/ctt}

\subsection{Encoding Type Theories in Dedukti}


Encoding a logic $L$ in Dedukti usually consists in introducing a
Dedukti theory (i.e. a set of constants and rewrite rules) $D(L)$, and
a mapping $[\![\,]\!]_L$ from $L$-formulae to Dedukti types and from $L$-proofs
to terms of type corresponding to the formula they prove.


In the case of Type Theory, one introduces a Dedukti type for ``codes
of types'', and a decoding function that assigns a Dedukti type to
each of these codes. Then, one introduces one constant for each type
constructor, and constants for introduction and elimination rules.

The basic property of this encoding is that it must be well-typed, in
the sense that a well-formed type must be translated to a well-typed
Dedukti term. More specifically this requires that definitionally
equal terms must be translated to convertible Dedukti terms. In other
terms, we expect that the definitional equality can be expressed as
rewrite rules.

As we have explained above, Cubical is a type theory which
definitional equality includes the equational theory of a de Morgan
algebra. It is far from obvious that it can actually be encoded by the
Dedukti rewrite rules.

We prefer to investigate another approach, where part of the
conversion is mapped to a kind of propositional
equality. Unfortunately, we cannot express conversion as a
propositional equality of Cubical (for the same reason that some
notions in HoTT cannot be expressed by mere axioms).

Those remarks have led to the introduction of Two Level Type Theories~\cite{Danil}.


\subsection{Two Level Type Theories}

We recall that in Type Theory there are two notions of equality:
\begin{itemize}
\item the propositional equality, that represents the intended equality of the logic
\item the definitional equality, which in fact is a judgment, which represents objects that should be identified to ensure important properties of the judgments
\end{itemize}

Two-Layer Type Theories are a class of type theories. Their common point is that they are in fact made of two types theories (both copies of MLTT, with different additional axioms)

\begin{itemize}
    \item \textbf{The internal}: It represents the theory we want to study. It is often equipped with Univalence Axiom, which makes its equality different than the usual equality, and incompatible with axioms like Uniqueness of Identity Proofs (UIP, or K) and functional extensionality (FunExt).
    
    \item \textbf{The external}: This theory will act as a sort of 'meta-theory' of the internal. We will use its propositional equality as an intermediate equality between the definitional ones (there are two definitional equalities here, the internal and the external), and the propositional equality of the internal theory.
    It has additional axioms (UIP \& FunExt) to make its propositional equality not slightly weaker but as powerful as the usual one.
\end{itemize}

        


%% file: sub/theory/ctt.tex


\def\intv{\mathbb{I}}
\def\face{\mathbb{F}}

The intuition behind Cubical is to follow the definition of paths as continuous functions from interval $[0;1]$ to the points of the topological space.

Cubical is a type theory introducing an interval pretype $\intv$.\footnote{$\intv$ is only a \emph{pretype}, as it does not enjoy all properties of types: we should not identify $0$ and $1$ although they are connected by a path.} 

Having an interval variable $i$ in the context, a judgement $i:\intv \vdash t : A$ represents a point $t$ parameterized by the interval, hence a path in $A$ (see Fig.~\ref{fig:square}, left).
From that, one can define a type \textsf{Path} and a path constructor in a way similar to $\lambda$-abstraction. It is also possible to apply an expression to path to get back a point of $A$.
$$\frac{\Gamma\vdash A ~\textrm{type}}{\Gamma \th \langle i \rangle t : \textsf{Path}~A~t(i0)~t(i1)}
\qquad
\frac{\Gamma\vdash p: \textsf{Path}~A~x~y \quad \Gamma\vdash e : \intv}{\Gamma\vdash p\,e : A}
$$

Formally, the interval $\intv$ is defined in a synthetic way: as the free De Morgan algebra on $i,j,k \cdots$.
Expanding the definition, this means its terms are elements of the following form :
$0 \mid 1 \mid i \mid 1 - r \mid r \wedge s \mid r \vee s $, with $\wedge$ representing the inf and $\vee$ representing the sup of the elements.

\begin{figure}
    \centering
\begin{tikzpicture}
\node (a) at (0,0)
{
\begin{tikzpicture}[scale=1.3]

\coordinate (O) at (0,0) ;
\coordinate (E) at (0,1) ;

\coordinate (fO) at (2.5,0.5) ;
\coordinate (fi) at (3,0.75) ;
\coordinate (fj) at (2,1) ;
\coordinate (fE) at (2.75,1.5) ;

\draw (O) node[below left]{$0$} node{$\bullet$};
\draw (E) node[above left]{$1$} node{$\bullet$};
\draw [->, cyan] (O) -- (E) node[midway, left]{$i$} ; 

\draw ($(O) - (0,0.5)$) node{$\mathbb{I}$};
\draw ($(O) + (2.5,-0.5)$) node{Type $A$};

\draw [dashed, ->] (O) to[bend right] (fO) ; 
\draw [dashed, ->] (E) to[bend right] (fE) ;

\draw [cyan] plot [smooth, tension=1] coordinates {(fO) ($(fO) + (0.125,0.3)$) ($(fO) + (-0.125, 0.6)$) (fE)};

\draw plot [smooth cycle] coordinates {($ (fO) - (0,0.45) $) ($ (fO) - (0,0.49) + (0.6,0) $) ($ (fi) + (0.355,0) $) ($ (fi) + (0.255,0.2) $)  ($ (fE) + (0.25,0.25) $) ($ (fj) - (0.25,0.125) $)};

\end{tikzpicture}
};
\node(b) at (7,0)
{
\begin{tikzpicture}[scale=1.5]

\coordinate (O) at (0,0) ;
\coordinate (i) at (1,0) ;
\coordinate (j) at (0,1) ;
\coordinate (E) at (1,1) ;

\coordinate (fO) at (2.5,0.5) ;
\coordinate (fi) at (3,0.75) ;
\coordinate (fj) at (2,1) ;
\coordinate (fE) at (2.75,1.5) ;

\draw ($(O) + (0.5,-0.5)$) node{$\mathbb{I}^2$};
\draw ($(O) + (2.8,-0.5)$) node{Type $A$};

\draw [->] (O) -- (i) node[right]{$i$} node[midway, below] {$j = 0$} ; 
\draw [->] (O) -- (j) node[left]{$j$} node[midway] {$i = 0$} ;
\draw [dotted, ->] (i) -- (E) node[midway] {$i = 1$}; 
\draw [dotted, ->] (j) -- (E) node[midway, above] {$j = 1$};


\draw [dotted, ->] (fO) -- (fi) ; 
\draw [dotted, ->] (fO) -- (fj) ;
\draw [dotted, ->] (fi) -- (fE) ; 
\draw [dotted, ->] (fj) -- (fE) ;


\draw [dashed, ->] (O) to[bend right] (fO) ; 
\draw [dashed, ->] (i) to[bend right] (fi) ;
\draw [dashed, ->] (j) to[bend left] (fj) ;
\draw [dashed, ->] (E) to[bend left] (fE) ;

\draw plot [smooth cycle] coordinates {($ (fO) - (0,0.45) $) ($ (fO) - (0,0.49) + (0.6,0) $) ($ (fi) + (0.355,0) $) ($ (fi) + (0.255,0.2) $)  ($ (fE) + (0.25,0.25) $) ($ (fj) - (0.25,0.125) $)};

\draw[pattern=north west lines, pattern color=gray] (O) rectangle (E);

\end{tikzpicture}
};
\end{tikzpicture}
    \caption{Meaning of judgments $i:\intv\vdash t:A$ and $i,j:\intv \vdash t:A$}
    \label{fig:square}
\end{figure}

Now if we have two interval variables, a judgment $i,j:\intv \vdash t:A$
means $t$ is a square in $A$, as illustrated by Fig.~\ref{fig:square}. Having $n$ interval variable leads to a $n$-dimensional cube in $A$, hence the name of Cubical Type Theory.

Some primitives of Cubical refer to expressions that may be defined only on a sub-polyhedra.
To do that, we first describe the cube in a synthetic way like we did with the interval.
A pretype $\face$ for the faces of the cube are defined with the following grammar:

\begin{itemize}
    \item $1$, the entire cube.
    \item $0$, the empty face.
    \item $i = 0$/$i = 1$ the face where $i = 0$/$i = 1$
    \item $f_1 \wedge f_2$, the intersection of the faces $f_1$, $f_2$
    \item $f_1 \vee f_2$, the union of the faces $f_1$, $f_2$
\end{itemize}

Contexts may also contain a face to restrict judgments to a sub-polyhedra. For instance, the judgment
$i,j:\intv;\,i=0 \vee j=1 \vdash t:A$ represents the left and top edges of the square in Fig.~\ref{fig:square}.

In the general case, type-checking in Cubical features a decision
procedure for the inclusion of faces. 
This is the main challenge in encoding Cubical in Dedukti.


%% file: sub/theory/2ltt.tex
In this section, we define Two Layers Type Theories by giving their encoding in Dedukti.

For a more comprehensive definition of 2LTT, we refer to~\cite{Danil}, although we had to adapt the definition as they were defined semantically on some specific points.

2LTTs are basically two copies of Martin L\"of's Type Theory: one internal layer and an external one.
In order to avoid the complexity of having the notion of type and later introduce each universe as a subclass (as is usual in MLTT), we parameterize the notion of type by \emph{levels}, that identify each universe, following Assaf (section 8.3 of~\cite{assaf}). We made the minimal assumptions of those levels, by just assuming a function \lst+lsuc+ such that universe $l$ belongs to level \lst+lsuc+\,$l$. The Dedukti declarations for that are:  
\begin{lstlisting}
Lev : Type.
lsuc : Lev -> Lev.
\end{lstlisting}

We can now introduce two codes of types: one for the internal layer (\lst+T+) and one for the external one (\lst+xT+), and their corresponding decoding functions \lst+eps+ and \lst+xeps+. Let us point out that we have chosen a shallow embedding where 2LTT contexts are identified with Dedukti contexts. So codes of types are not explicitely parameterized by a notion of context.


        
        
        
        
\begin{lstlisting}
T : Lev -> Type.                   xT : Lev -> Type.
def eps : i : Lev -> T i -> Type.  def xeps : i:Lev -> T i -> Type.
\end{lstlisting}

In this encoding, an internal type $A$ at level $l$ is a term \lst+A : T l+, an element $t$ of that type $A$ is a term \lst+t : eps l A+. \lst+T+, and likewise for external types.

The inclusion of a universe $l$ in the bigger universe is taken care by first introducing a code in \lst+t l : T (lsuc l)+, and a rewrite rule to assert that \lst+t l+ decodes to \lst+T l+:
\begin{lstlisting}
t : i : Lev -> T (lsuc i).       xt : i : Lev -> xT (lsuc i).
[i] eps (lsuc i) (t i) --> T i.  [i] xeps (lsuc i) (xt i) --> xT i.
\end{lstlisting}

It remains to lift each type of level $l$ as a type of level \lst+lsuc+\,$l$. We omit the external lift which is defined similarly.
\begin{lstlisting}
lUp :  i : Lev -> a : T i -> T (lsuc i).
[i, a] eps (lsuc i) (lUp i a) --> eps i a.
\end{lstlisting}
The above rewrite rule relate codetypes at level $l$ and their counterpart at level \lst+lsuc+\,$l$ in a very strong way: they decode to the same type, which means they have the \emph{same} inhabitants.

We then implemented the usual primitive types of MLTT to populate the universes. As an example, internal dependent pairs are declared by introducing a constant \lst+Sig+ for the typecode, \lst+pair+ is the introduction rules, and \lst+p1+ and \lst+p2+ are the projections:
\begin{lstlisting}
Sig : i : Lev -> A : T i -> (eps i A -> T i) -> T i.
def tSig := (i : Lev => A : T i => B : (eps i A -> T i) 
  => eps i (Sig i A B)).

def pair : i : Lev -> A : T i -> B : (eps i A -> T i) -> 
    a : eps i A -> b : eps i (B a) -> tSig i A B.
def p1 : i : Lev -> A : T i -> B : (eps i A -> T i) -> 
    p : tSig i A B -> eps i A.
def p2 : i : Lev -> A : T i -> B : (eps i A -> T i) ->
    p : tSig i A B -> eps i (B (p1 i A B p)).
    
[i,A,B,a,b] p1 i A B (pair i A B a b) --> a.
[i,A,B,a,b] p2 i A B (pair i A B a b) --> b.
\end{lstlisting}

Actually, we define the following types, both at the internal and external layer, and at each level:

{
\begin{center}
\begin{tabular}{|c|c|}
    \hline
    Internal   & External     \\
    \hline
   \lst+False i : T i+ &
   \lst+xFalse i : xT i+ \\
   \lst+True i : T i+  &
   \lst+xTrue i : xT i+  \\
   \lst+Nat i : T i+   &
   \lst+xNat i : xT i+   \\
   \lst+Pi i (A:T i)(B:x:eps i A->T i) : T i+ &
   \lst+xPi i (A:xT i)(B:x:xeps i A->xT i) : xT i+ \\
   \lst+Sig i (A:T i)(B:x:eps i A->T i):T i+   &
   \lst+xSig i (A:xT i)(B:x:xeps i A->xT i):xT i+   \\
   \lst+Sum i (A:T i) (B:T i) : T i+   &
   \lst+xSum i (A:xT i) (B:xT i) : xT i+   \\
   \lst+Eq i (A:T i)(x:eps i A)(y:eps i A):T i+    &
   \lst+xEq i (A:xT i)(x:xeps i A)(y:xeps i A):xT i+    \\
   \hline
\end{tabular}
\end{center}
}

So far, 2LTTs feature two copies of MLTT, each one totally independent from the other. In order to include the internal layer into the external one, 2LTTs feature a coercion function \lst+c+ that assigns an external to each internal type.
\begin{lstlisting}
def c  :  i : Lev -> T i -> xT i.
\end{lstlisting}
    
In~\cite{Danil}, the coercion was defined in semantical terms. Here, coercion is such each internal type $A$ is \emph{isomorphic} to \lst+c+$(A)$. By lifting internal types into the external world, the coercion allows us to see the internal world as a sort of sub-world of the external world. This allows to express properties of the internal world using the external equality.

This isomorphism can be encoded in different ways, from the most general to the most specific:
\begin{itemize}
    \item Assuming the existence of functions between \lst+eps+$(A)$ and \lst+xeps+$($\lst+c+$(A))$, which are inverse one of each other, propositionally.
    \item Assuming the existence of functions between \lst+eps+$(A)$ and \lst+xeps+$($\lst+c+$(A))$, which are inverse one of each other, definitionally.
    \item Assuming that both (code)types decode to the same type.
\end{itemize}
The third option is similar to the one chosen for the universe inclusion, but the goal of 2LTTs is to be as general as possible, so we would better avoid it. However, the first one would probably need a coherence condition, and would make the system harder to use. For these reasons we have chosen the second option:
\begin{lstlisting}
def isoUp :   i : Lev -> A : T i -> eps i A -> tc i A. 
def isoDown : i : Lev -> A : T i -> tc i A  -> eps i A.
[i, A, a] isoDown i A (isoUp i A a) --> a.
[i, A, a] isoUp i A (isoDown i A a) --> a.
\end{lstlisting}

As stated before, 2LTTs are a class of type theories. They are a sort of 'à la carte' type theory where one can add additional axioms concerning coercion to tune it the way one wants it to be.

Like the definition of coercion, most of these axioms had a semantic definition. Here is a list of the ones (as introduced in~\cite{Danil}) we were able to express in a syntactic manner:

\begin{itemize}
    \item Coercion can be required to be injective:
    
\begin{lstlisting}
(; < T1 > ;)
def T1 : l : Lev -> A : T l -> B : T l ->
         p : xtTEq l (c l A) (c l B) -> tTEq l A B.
\end{lstlisting}
    where \lst+xtTEq+ and \lst+tTEq+ are shorthands for the types of elements of respectively external and internal equality
    
    \item The repletion axiom: an external type isomorphic to a $c(A)$ also has an antecedent by $c$.
\begin{lstlisting}
(; < T3 > ;)
repletion : l : Lev -> A : xT l -> B : T l ->
            p : xtTEq l A (c l B) -> T l.
[l, A, B, e] c l (repletion l A B e) --> A.
\end{lstlisting}
        
    \item Integers are a fairly simple type with simple rules. One would expect $c(\mathbb{N})$ and x$\mathbb{N}$ to be isomorphic, but it is actually not the case for technical reasons. There is only a morphism x$\mathbb{N} \rightarrow c(\mathbb{N})$.
    
    A possible additional axiom is to make this morphism an isomorphism definitionally.
    
    The union, eq and false types can have similar additional axioms, while in the case of the product, sum and 1 types, the isomorphism is already there. 
        
    \item One can also make these types isomorphic through that isomorphism not definitionally (ie by rewriting), but only up to external equality.
    
    \item One can make all these isomorphisms (including the ones for pi, sig, and 1, with example code below) equalities instead of just isomorphisms (cf option 3 for the coercion).
    
\begin{lstlisting}
(; < T2 >  Primitive Isomorphisms c A ~ xA become equality ;)
[l] c l True --> xTrue l.
[l, A, B] c l (Pi l A B) --> xPi l (c l A) (clift l A B). 
[l, A, B] c l (Sig l A B) --> xSig l (c l A) (clift l A B).
\end{lstlisting}

\end{itemize}

We implemented all the above axioms in Dedukti.

Interestingly, there was in the list of axioms that couldn't be implemented an axiom called $(A5)$, which requires that external equality validates the reflection rule (that is, externally equal types are considered definitionally equal). This could not be implemented in Dedukti. In it was possible, then we would be able to encode Cubical in Dedukti without resorting to 2LTTs.

We also tested the usability of this encoding by formulating the axiom of univalence. Here, for the sake of brevity, we only give the weaker form which states that the type $A=B$ (where $A$ and $B$ are types of level $l$) is weakly equivalent to $A\approx B$, the type of weak equivalence between $A$ and $B$:
\begin{lstlisting}
WeakUnivalence : l : Lev -> A : T l -> B : T l ->
  eps (lsuc l) (Equiv (lsuc l) (TEq l A B) (lUp l (Equiv l A B))).
\end{lstlisting}
Note that $A=B$ actually belongs to level \lst+lsuc+\,$l$, hence the need to lift $A\approx B$ one universe up. Also, the notion of weak equivalence occurs twice but at different levels. This remark is the reason that made us opt for a presentation of type theories with a hierarchy of universes

%% file: sub/implem/translation.tex

We define a straightforward translation, with every type/term associated to the one with the same name in Dedukti, and the convention that variables share the same name in 2LTT and Dedukti:

\begin{itemize}
    \item $\Trad{l}{x} =$ \lst+x+ 
    \item $\Trad{l}{\Sum{}{A}{B}} =$ \lst+Sig l + $\Trad{l}{A}$ \lst+(x: eps l +$\Trad{l}{A}$\lst+ => +$\Trad{l}{B(x)}$\lst+)+
    \item$\Trad{l}{\pair{}{A}{B}{a}{b}} =$ \lst+pair l +$\Trad{l}{A}$ \lst+(x: eps l +$\Trad{l}{A}$\lst+ => +$\Trad{l}{B(x)}$\lst+)+ $\Trad{l}{a}$ $\Trad{l}{b}$
    \item $\cdots$
\end{itemize}

We also define the translation of context :
\begin{center}
  $\Trad{}{x_1 :_{l_1} A_1, \cdots, x_n :_{l_n} A_n} =$ \lst+x1: eps + $\Trad{l_1}{A_1}$, ..., \lst+xn: eps + $\Trad{l_n}{A_n}$.
\end{center}

As stated before when we first talked about how 2LTTs were implemented, encoding has the particularity that it encodes types, not into Dedukti types, but into type codes, that is elements of type \lst+T+\,$l$ for internal types, and \lst+xT+\,$l$ for external types. It encodes terms into terms of type the 'realization' of $\Trad{l}{A}$, that is \lst+eps +$l~\Trad{l}{A}$, and similarly for the external layer.

%% file: sub/implem/soundness.tex

There are two important properties that the translation is expected to have:
        
\begin{description}
    \item[Soundness:] this means that the translation defined above preserves typability, and hence provability too. It also means that our encoding is powerful enough to prove everything that could be proven in the theory of 2LTTs.
    \item[Conservativity:] this would mean that any property provable in the encoding can be proved in the thoery of 2LTTs, in other words that our encoding is not too powerful. From this we would be able to use Dedukti as a 2LTT type-checker.
\end{description}

While soundness is quite straightforward to prove (since all of the definitional equality of 2LTTs could be encoded by rewrite rules), conservativity is quite hard to prove.

%




The soundness property consists of 9 statements, one for each judgment kind:

\begin{Theorem}
    
    \begin{itemize}
    
        \item If $\Gamma $ 2LTT context, then $\Trad{l}{\Gamma}$ Dedukti context.
        
        
        \item If $\Gamma \th_{2L} A$ type l, then $\Trad{l}{\Gamma} \th_{Dk} \Trad{l}{A} : $ \lst+T l+
        
        \item If $\Gamma \th_{2L} A = A'$ type l, then $\Trad{l}{\Gamma} \th_{Dk} \Trad{l}{A} \leftrightarrow^* \Trad{l}{A'} : $ \lst+T l+
        
        \item If $\Gamma \th_{2L} A$ type l, $\Gamma \th_{2L} t : A$, then $\Trad{l}{\Gamma} \th_{Dk} \Trad{l}{t} : $ \lst+eps l +$\Trad{l}{A}$
        
        \item If $\Gamma \th_{2L} A$ type l, $\Gamma \th_{2L} t = t' : A$ then $\Trad{l}{\Gamma} \th_{Dk} \Trad{l}{t} \leftrightarrow^* \Trad{l}{t'} : $ \lst+eps l + $\Trad{l}{A}$
        
        
        \item If $\Gamma \th_{2L} A$ xtype l, then $\Trad{l}{\Gamma} \th_{Dk} \Trad{l}{A} : $ \lst+xT l+
        
        \item If $\Gamma \th_{2L} A = A'$ xtype l, then $\Trad{l}{\Gamma} \th_{Dk} \Trad{l}{A} \leftrightarrow^* \Trad{l}{A'} : $ \lst+xT l+
        
        \item If $\Gamma \th_{2L} A$ xtype l, $\Gamma \th_{2L} t : A$, then $\Trad{l}{\Gamma} \th_{Dk} \Trad{l}{t}$ \lst+: xeps l +$\Trad{l}{A}$
        
        \item If $\Gamma \th_{2L} A$ xtype l,$\Gamma \th_{2L} t = t' : A$ then $\Trad{l}{\Gamma} \th_{Dk} \Trad{l}{t} \leftrightarrow^* \Trad{l}{t'}$ \lst+: xeps l + $\Trad{l}{A}$
    \end{itemize}
        
\end{Theorem}
The proof is by mutual induction on the judgments.

Conservativity is a much harder problem, and we have not proven it yet. However, we make the following conjecture a conservativity result. Every Dedukti proof which context and type are in the image of the translation correspond to a 2LTT derivation: 
\begin{Conjecture}
Given a Dedukti judgment
$$ x_1:T_1,\,\ldots, x_n:T_n \th_{Dk} t : A$$
where all $T_i$s are of the form \lst+eps+\,$l_i\,\Trad{l_i}{U_i}$ or \lst+xeps+\,$l_i\,\Trad{l_i}{U_i}$ and $A$ is convertible to a term of the form \lst+eps+\,$l\,\Trad{l}{B}$ or \lst+xeps+\,$l\,\Trad{l}{B}$, then there exists a 2LTT terms $u$ such that 
$$x_1 : U_1,\,\ldots,\,x_n:U_n \th_{2L} u : B$$ 
\end{Conjecture}
This obviously cannot hold if the logic of Dedukti is inconsistent (unless 2LTTs are themselves inconsistent). The idea of the proof is to translate only proofs in normal form, and assume strong normalization of the reduction rules of Dedukti. Another source of inspiration is~\cite{assaf-completeness}, where conservativity is proven for an encoding of Pure Type Systems.


%% file: sub/explanations/clttare2ltt.tex
This section introduces a partial encoding of Cubical as an extension of the 2LTT Dedukti theory. We focused of the main primitive of Cubical: composition. As we have already mentionned, the typing rule of composition is probably beyond the capabilities of Dedukti's rewrite rules, and we expect 2LTTs to be a trade-off where expressivity is recovered at the cost of building parts of definitional equality derivations by hand. We try to express the largest fragment by rewrite rules, and the rest will be encoded in the external layer.

We do not consider the primitives related to glueing, which is the main feature that makes univalence provable in Cubical.

When viewing Cubical as an instance of a 2LTT, the leading idea is that the internal layer contains the object theory (Cubical) while the external layer is that of the meta-theory. More concretely, the internal layer contains the types of Cubical, while the external layer contains the pretypes ($\intv$, and $\face$) and the judgments of Cubical.

 

We first introduce a level \lst+cL+ for the primitive pretypes of Cubical: $\intv$ and $\face$ which are declared as external types.
\begin{lstlisting}
cL : Lev.              def T := xT cL.        def ceps := xeps cL.
def cEq := xEq cL. (; and all types at level cL with prefix c ;)
\end{lstlisting}

Symbol \lst+cEq+ is used to express convertibility of preobjects. For the sake of conciseness we also define a symbol to represent convertibility of objects, using the coercion to lift internal objects to the external layer: 
\begin{lstlisting}
def CubicalEq (l : Lev) (A : T l) (a : eps l A) (b : eps l A) :=
      xEq l (c l A) (isoUp l A a) (isoUp l A b).
\end{lstlisting}

In order to interpret the conversion rule, we need more interaction between both layers, by allowing elimination of an external equality at level \lst+cL+ towards an internal type:
\begin{lstlisting}
def CubicalJ :
  l:Lev -> A:cT -> x:ceps A -> P: (y:ceps A->cEq A x y->T l) ->
  eps l (P x (crefl A x)) ->
  y:ceps A -> e:cEq A x y -> eps l (P y e). 
\end{lstlisting}

\subsection{The interval pretype $\intv$}

Implementing the grammar of intervals and faces needs care, because the type-checking of Dedukti requires confluence of the set of rewrite rules. An algebra $(A,\vee,\wedge,0,1,\neg)$  is a De Morgan algebra if $\vee$ and $\wedge$ are associative and commutative and
$$\begin{array}{c}
x\wedge x = x \qquad x \wedge 0 = 0 \qquad x \wedge 1 = x \\
x \wedge (y \vee z) = (x \wedge y) \vee (x \wedge z) \qquad \neg(x\wedge y) = \neg x \vee \neg y \qquad \neg \neg x = x
\end{array}$$
The dual laws are derivable from these. In the currently distributed version of Dedukti, commutativity cannot be added as a rewrite rule, and idempotence being non-linear may break confluence. Zero, neutral, involution and De Morgan laws are straightforward. Associativity can be oriented in an arbitrary direction. Regarding distributivity, we cannot have a law and the dual one (neither the left nor the right one) or normalization is lost. Having the left and right laws at the same time creates a critical pair that cannot be closed without commutativity. So, we can have at most one of the four. Considering that distributivity is probably used only in very few cases, we decided to implement none as a rewrite rule. The rules that cannot be expressed as rewrite rules are thus stated as external equations.
\begin{lstlisting}
I : cT.          0 : ceps I.        1 : ceps I.
def Imin : ceps I -> ceps I -> ceps I.
def Imax : ceps I -> ceps I -> ceps I.
def sym : ceps I -> ceps I.
(; rewrite rules, completed by symmetry ;)
[i] Imin 0 i --> 0  [i] Imin i 0 --> 0.
[i] Imin 1 i --> i [i] Imin i 1 --> i.
[i] Imax 0 i --> i  [i] Imax i 1 --> 0.
[i] Imax 1 i --> 1 [i] Imax i 1 --> 1.
[] sym 0 --> 1  [] sym 1 --> 0.
[i,j] sym (Imin i j) --> Imax (sym i) (sym j).
[i,j] sym (Imax i j) --> Imin (sym i) (sym j).
[i] sym (sym i) --> i.
[i,j,k] Imin (Imin i j) k --> Imin i (Imin j k).
[i,j,k] Imax (Imax i j) k --> Imax i (Imax j k).
(; properties expressed as external equations;
   Imin laws derived by duality ;)
Imax_idem : i:ceps I -> cEq (Imax i i) i.
Imax_comm : i:ceps I -> j:ceps I -> cEq I (Imax i j) (Imax j i).
Imax_dist : i : ceps I -> j : ceps I -> k : ceps I ->
   cEq (Imax (Imin i j) k) (Imin (Imax i k) (Imax j k).
\end{lstlisting}

\subsection{Paths}

We then define the type of paths together with its introduction and elimination rules. Since paths are types in Cubical, they are encoded in the internal layer.
\begin{lstlisting}
def Path :   A : cT -> u : ceps A -> v : ceps A -> cT.
def lam : A : cT -> p : (ceps I -> ceps A) ->
     ceps (Path A (p 0) (p 1)).
def app : A : cT -> u : ceps A -> v : ceps A ->
     ceps (Path A u v) -> ceps I -> ceps A.
(; Computational rules ;) 
[A,u,v,p] app A u v (lam A f) e --> f e.   (; beta ;)
[A,u,v,p] app A u v p 0 --> u  [A,u,v,p] app A u v p 1.     
\end{lstlisting}
In the definition of Cubical, the last two definitional equalities above implement the rules
$$\frac{\Gamma\vdash p : \textsf{Path}~A~u~v}{\Gamma\vdash p~0 = u \,:A\qquad \Gamma\vdash p~1 = v \,:A}$$
which requires typing information about path $p$. This could be a problem since the conversion (and rewrite rules) of Dedukti is applied on terms without any typing information. Fortunately, in our encoding application is annotated with all of the information needed.

We also note that at this point paths are not related to the internal equality \lst+Eq+.


It remains to express the key primitive of Cubical: composition. The typing rule involves the notions of interval variable (which are just external variables of type $\intv$) and face ($\face$).

\subsection{Faces}

Let us first define faces, following the explanations in the introduction.
\begin{lstlisting}
F : cT. (; Type of Faces;)
0f : ceps F. (; Empty face ;)
1f : ceps F. (; Whole cube ;)
def eq0 : ceps I -> ceps F. (; eq0 i is the face i = 0 ;)
def eq1 : ceps I -> ceps F. (; eq1 i is the face i = 1 ;)
def Fmin: ceps F -> ceps F -> ceps F. (; Intersection of faces ;)
def Fmax: ceps F -> ceps F -> ceps F. (; Union of faces ;)
(; Rewrite rules and equations (problem similar to the interval);)
[f] Fmin 0f f --> 0f.
...
Fdiscr : i : ceps I -> cEq F (Fmin (eq0 i) (eq1 i)) 0f.
\end{lstlisting}
We do not give details of how the algebraic properties of faces are turned into either rewrite rules or an equation. We only give the crucial property \lst+Fdiscr+ that there is no intersection between the opposite faces of a cube.

An important remark is that this does not exactly follow the syntax of the faces of Cubical, since the face $(i=1)$ requires $i$ to be a \emph{variable}, which cannot be enforced in our shallow embedding. In Cubical when an interval variable $i$ is substituted by an interval expression $e$, in a face $(i=1)$, it is replaced following the rules (and similarly for $i=0$):
$$\begin{array}{c}
(i=1)[i/0] = 0 \qquad
(i=1)[i/1] = 1 \qquad
(i=1)[i/1-e] = (i=0)[i/e]\\
(i=1)[i/e_1\vee e_2] = (i=1)[i/e_1] \vee (i=1)[i/e_2] \\
(i=1)[i/e_1\wedge e_2] = (i=1)[i/e_1] \wedge (i=1)[i/e_2]
\end{array}$$This is quite naturally expressed by rewrite rules (straightfowardly adapted to \lst+eq0+):
\begin{lstlisting}
[] eq1 0 --> 0f      [] eq1 1 --> 1f   [e] eq1 (sym e) --> eq0 e.
[i,j] eq1 (Imax i j) --> Fmax (eq1 i) (eq1 j).
[i,j] eq1 (Imin i j) --> Fmin (eq1 i) (eq1 j).
\end{lstlisting}

Given a context $\Gamma$ and a face $\phi$ of $\Gamma$, the context $\Gamma,\phi$ is a restriction of context $\Gamma$ where the interval variables must belong to $\phi$. Since we use a shallow embedding of context, we need to represent a face as an external type (actually a proposition) of witnesses that the interval variable belong to $\phi$. So, the Dedukti type \lst+F+ above is a code of types with a decoding function \lst+faceType+.

The definition one would like to make would thus be something along the lines of:
\begin{lstlisting}
(; First attempt ;)
def faceType : ceps F -> cT.
[]     faceType 0f         --> cFalse.
[]     faceType 1f         --> cTrue.
[a]    faceType (eq1 a)    --> cEq I 1 a.
[a]    faceType (eq0 a)    --> cEq I 0 a.
[a, b] faceType (Fmax a b) --> cSum (faceType a) (faceType b).
[a, b] faceType (Fmin a b) --> cSig (faceType a) (_=>faceType b).
\end{lstlisting}
where we see that intersection (resp. union) is represented by the cartesian product (resp. sum), and the base constraint $(i=0)$ by an external equality.

But this definition breaks confluence. Here is an example of critical pair:
\begin{lstlisting}
faceType (Fmin 1f 1f) --> cSig cTrue (_=> cTrue)
faceType (Fmin 1f 1f) --> faceType 1f --> cTrue
\end{lstlisting}
If we try to recover confluence by closing this critical pair, the types \lst+cTrue+ and 
\lst+cSig cTrue+ \lst+(_=> cTrue)+ become convertible and hence allow to apply a projection to an inhabitant of \lst+cTrue+. This destroys many good properties (e.g. canonicity) of the external layer, and this may lead to have erroneous Cubical proofs accepted by the encoding.

As a workaround, we decided to not have rewrite rules associated to \lst+faceType+, but rather have an \emph{isomorphism} between \lst+faceType+\,$\phi$ and its intended type. We will also need that those types are actually propositions (i.e. all inhabitants must be equal). This is the case for faces $0$, $1$ and $\phi_1 \wedge \phi_2$. This holds also for $(i=1)$ and $(i=1$) if the external layer enjoys Uniqueness of Identity Proofs (or axiom K). But having $\phi_1 \vee \phi_2$ isomorphic to a disjoint sum is a problem because the latter type may not be a proposition. We need a new external type, for instance a truncated sum. This type as the same introduction rules similar to disjoint sum, but the elimination rule requires a coherence condition (that will be given in the definition \lst+TermSys+ below).






%% file: sub/implem/cttsys.tex
We now focus on the Cubical notion of \emph{system}, which allows to define functions whose value depends on where we are on the cube. A system is an expression of the form $[\phi_1 \rightarrow a \mid \phi_2 \rightarrow b]$, which evaluates to $a$ on $\phi_1$, and $b$ on $\phi_2$. This expression is defined on $\phi_1 \vee \phi_2$ which is called the extent of that system. Obviously, this makes sense only when $a$ and $b$ coincide on $\phi_1\wedge \phi_2$ w.r.t. definitional equality. In Cubical, systems are valid terms only if their extent is equal to $1f$.

In our encoding, a system of type $A$ and extent $\phi$, is a term of type \lst+ceps(faceType phi)+ \lst+-> eps l A+. The embedding into terms of type $A$ is done by applying a proof that the current context is on face $\phi$. So, a (binary) partial system is build with the following symbol:
\begin{lstlisting}
def TermSys : l : Lev -> f1 : ceps F -> f2 : ceps F ->
   tau : T l ->
   A1 : (ceps(faceType f1) -> eps l tau) ->
   A2 : (ceps(faceType f2) -> eps l tau) ->
   coh : (e : ceps (faceType (Fmin f1 f2)) -> 
          tCubicalEq l tau (A1 (fp1 f1 f2 e)) (A2 (fp2 f1 f2 e))) ->
   ceps (faceType (Fmax f1 f2)) ->  eps l tau.
\end{lstlisting}

which implements the rule
$$\frac{\Gamma\th A~\text{type}\qquad 
 \Gamma,\phi_1\th a_1 : A \qquad \Gamma,\phi_2\th a_2:A \qquad
 \Gamma,\phi_1\wedge\phi_2\th a_1 = a_2 : A}%
{\Gamma,\phi_1\vee\phi_2\th [\phi_1\rightarrow a_1 \mid \phi_2\rightarrow a_2] : A}$$

Actually, we implemented a more general rule where the first premisse is $\Gamma,\phi_1\vee\phi_2\th A~\text{type}$, but this adds a lot of technicalities on the handling of face witnesses. We shall not give all the details here, but making those witnesses irrelevent (either definitionaly or propositionaly) is necessary.


This condition and the fact that the theory uses dependent types makes the use of systems in practice really complex, especially when more than two branches are involved since one has to check multiple side conditions.


%

%% file: sub/implem/cttfill.tex
The formal definition of composition is

$$\frac{
 \Gamma \th \phi : \face \qquad
 \Gamma, i : \intv \th A~\text{type} \qquad
 \Gamma, \phi, i : \intv \th u : A \qquad
 \Gamma \th a_0 : \Inst{A(i0)}{\phi \rightarrow u(i0)}}%
{\Gamma \th \text{comp}^i~A~[\phi \rightarrow u]~a_0~:~\Inst{A(i1)}{\phi \rightarrow u(i1)}}
$$
where the notation $\Gamma\th t:\Inst{A}{\phi\rightarrow u}$ is a shorthand for
$\Gamma\th t:A$ and $\Gamma,\phi\th u=u:A$, in other words $t$ as type $A$ and is definitionally equal to $u$ on face $\phi$. This cannot be expressed via rewriting. So we used external equality both for the premisse and the conclusion, which is split into two symbols:

\begin{lstlisting}
def primCompTerm : l : Lev -> phi : ceps F -> A : (ceps I -> T l)
    -> u : (ceps (faceType phi) -> i : ceps I -> eps l (A i))
    -> a0 : eps l (A 0) 
    -> (e : ceps (faceType phi)-> tCubicalEq l (A 0) a0 (u e 0))
    -> eps l (A 1).

def primCompEq : l : Lev -> phi : ceps F -> A : (ceps I -> T l)
     -> u : (ceps (faceDataType phi) -> i : ceps I -> eps l (A i))
    -> a0 : eps l (A 0)
    -> coh :(e:ceps(faceType phi -> tCubicalEq l (A 0) a0 (u e 0))
    -> e : ceps (faceType phi)
    -> tCubicalEq l (A 1) (primCompTerm l phi A u a0 coh) (u e 1).
\end{lstlisting}

Combining composition with systems, one can prove many important theorems such as transitivity of paths. To make an example use of our encoding as well as to test how usable it was in practice, we implemented the example of filling, which draws the line between a point $a_0$ and the composition $\text{comp}^i~A~[\phi \rightarrow u]~a_0$.:
$$ \text{fill}^i~A~[\phi \rightarrow u]~a_0~j~:~A(j)$$
Because of the verbosity of our encoding, making the proof by hand (\ie creating the fill term) was quite complicated in Dedukti. The situation would be better with the development version of Dedukti which features an interactive proof construction engine.

%% file: sub/implem/cttsoundness.tex
The soundness and conservativity of the encoding (using an translation function) is clearly a quite hard problem. Here we will only sketch how those results should be obtained.

The first step is to translate Cubical terms into a Dedukti well-typed terms in the theory of 2LTTs extended with the Cubical-specific piece of theory described above. This is quite difficult since parts of the definitional equality of Cubical is translated to external equalities of 2LTTs. 
Given two Cubical-convertible types $A$ and $B$, a term $M$ of types $A$ is also of type $B$, while in the 2LTT encoding, $\Trad{}{M}$ has type $\Trad{}{A}$, but getting a term of type $\Trad{}{B}$ requires the transport principle \lst+CubicalJ+. Formally, one would say that the translation domain is Cubical derivations rather than mere terms.

Proving the soundness of this translation raises the difficulty that the same term may be typed by different derivations (using conversion at different places), resulting in translated terms that may not be convertible. One important lemma is to show they are actually equal w.r.t. external equality. This is where it is important that the external equality satisfies axiom UIP and functional extensionality. We note that is problem is equivalent to the one of encoding extensional type theory into intensional type theory extended with the above two axioms, see~\cite{ExtensionalTT}

The conservativity proof follows the same idea as the one of 2LTTs.


%% file: concl.tex

%




We gave an encoding of 2LTTs as a Dedukti theory, and specialized it to an encoding of a subsystem of Cubical Type Theory (excluding glueing).

The 2LTT encoding was rather straightforward, once the typing rules are expressed syntactically. Definitional equality could be encoded as rewrite rules, which made the soundness proof rather easy. However, we consider 2LTTs as an important logical framework to express theories which definitional equality is very rich. Other theories could be expressed in our encoding. Beyond the HoTT family of formalisms, we may cite CoqMTU~\cite{CoqMTU}, an extension of type theory with a decidable first-order theory.

The Cubical encoding required much more care. There are two main reasons for this. Firstly, Cubical is based on an algebraic structure (De Morgan algebra) which properties cannot be encoded easily as rewrite rules (commutativity, distributivity, idempotence). The answer to this problem may be to extend the expressivity of the rewrite rules of Dedukti: rewriting modulo AC is being considered, but it is not clear if this work is useful for theories with more properties, like having a unit element. The second reason is that the definitional equality of Cubical includes a decision procedure. Rewrite rules cannot inspect the context, so it does seem possible to encode it without giving up the shallow embedding for faces.

    
\subsection*{Future work}
    
There are several directions that this work may take. Of course, one is to establish the soundness and conservativity results that we expect hold for our encodings.

We also plan to study how glueing can be added to the encoding. At first glance, the reduction rule associated to glue types cannot be expressed as rewrite rules, so there are probably several interesting problems to study there.

A more concrete concern would be to define the translation function from Cubical to our encoding. In this paper, we have done it by hand on several examples (like filling). It appeared to become very complex when systems are involved. Writing an elaboration function that fills the gaps would be the next step before instrumenting Cubical proof systems to produce Dedukti terms that could be rechecked.